\documentclass[
superscriptaddress,
amsmath,
amssymb,
aps,
twocolumn,
]{revtex4-2}
\usepackage{xcolor}
\usepackage{graphicx}
\usepackage{dcolumn}
\usepackage{bm}
\usepackage{hyperref}

\usepackage{varioref}
\usepackage{xr-hyper}
\makeatletter

\begin{document}

\title{Glassy phases of the Gaussian Core Model}

\author{Vittoria~Sposini}
\affiliation{Faculty of Physics, University of Vienna, Boltzmanngasse 5, 1090 Vienna, Austria}

\author{Christos~N.~Likos}
\affiliation{Faculty of Physics, University of Vienna, Boltzmanngasse 5, 1090 Vienna, Austria}

\author{Manuel~Camargo}
\affiliation{Facultad de Ciencias \& CICBA, Universidad Antonio Nari\~{n}o--Campus Farallones, Km 18 via Cali-Jamund\'i, 760030 Cali, Colombia}

\date{\today}

\begin{abstract}
We present results from molecular dynamics simulations exploring the supercooled dynamics of the Gaussian Core Model in the low- and intermediate-density regimes. In particular, we discuss the transition from the low-density hard-sphere-like glassy dynamics to the high-density one.  The dynamics at low densities is well described by the \emph{caging} mechanism, giving rise to intermittent dynamics. At high densities, the particles undergo a more continuous motion in which the concept of cage loses its meaning. We elaborate on the idea that these different supercooled dynamics are in fact the precursors of two different glass states. 
\end{abstract}


\maketitle

Soft colloids encompass a large variety of nano- or mesoscopic aggregates undergoing thermal motion in a solvent. In most cases, these are polymer-based assemblies of various architectures and connectivities, such as linear chains, stars and rings, dendrimers, block-copolymer micelles, but also cross-linked nano- or microgels, which feature a high degree of flexibility and deformability~\cite{vlassopoulos:cocis:2014}. The degree of softness is conveniently expressed by the ratio $E_{\mathrm{el}}/k_{\rm B}T$ of their elastic deformation energy upon small overlaps or indentations, $E_{\mathrm{el}}$, over the thermal energy $k_{\rm B}T$,  which can span several orders of magnitude~\cite{vlassopoulos:cocis:2014, riest:scirep:2015}.  
Their suspensions exhibit structural and dynamical anomalies~\cite{watzlawek:jpcm:1998, krekelberg:pre:2009, truskett:jcp:2011, pond:sm:2011}, which accompany rich types of thermodynamic behavior, such as reentrant melting~\cite{stillinger:jcp:1976, pamies:jcp:2009, berthier:pre:2010, gnan:natphys:2019, parisi:mm:2023}, and clustering~\cite{lenz:prl:2012, stiakakis:natcomm:2021}.
The deformability of these soft colloids manifests itself under flow as particle elongation, tumbling and shear-thinning~\cite{smith:science:1999, leduc:nature:1999, gerashchenko:prl:2006, ripoll:prl:2006, huang:mm:2010}.
Akin to hard colloids in the supercooled regime, soft colloids exhibit a sharp increase of the equilibrium relaxation time and heterogeneous dynamics~\cite{vlassopoulos:cocis:2014, foffi:prl:2003, mayer:natmat:2008, mattson:nature:2009}. 
However, at very high concentrations, soft colloids can reach a particular aging regime characterized by an intermittent release of internal stresses which coincides with the onset of an anomalous decrease in local order~\cite{philippe:pre:2018}. Colloidal softness has been also related to the microscopic origin leading to the validity of the Stokes-Einstein relation for degrees of metastability for which it normally breaks down in the case of hard colloidal and molecular systems~\cite{gupta:prl:2015}.

From a theoretical point of view, the effective interaction $v(r)$ between a pair of such colloids 
does not grow fast as $r \to 0$, so that the integral $\int_0^{\infty} v(r){\rm d}^3 r$ is finite. In some cases, these are effective interactions between the centers of mass of open, fractal, fully penetrable objects, for which $v(r)$ remains finite (free of divergence) even if the separation $r$ vanishes. Denoting with $\hat v(k)$ the Fourier transform of $v(r)$, such effective interactions are referred to as $Q^{+}$ potentials if $\hat v(k)$ is positive definite and as $Q^\pm$ potential if $\hat v(k)$ attains both positive and negative values, typically in an oscillatory fashion~\cite{likos:pre:2001}. Systems belonging to the $Q^{+}$ class undergo a reentrant fluid-crystal-fluid transition at low temperature and high density and they possess a maximum freezing temperature, beyond which no crystallization is possible whereas systems of the $Q^{\pm}$ class bond together forming cluster crystals, where each lattice site is occupied by several overlapping particles~\cite{likos:pre:2001, stiakakis:natcomm:2021}

\begin{figure}[tbhp]
    \centering
    \includegraphics[width=1\linewidth]{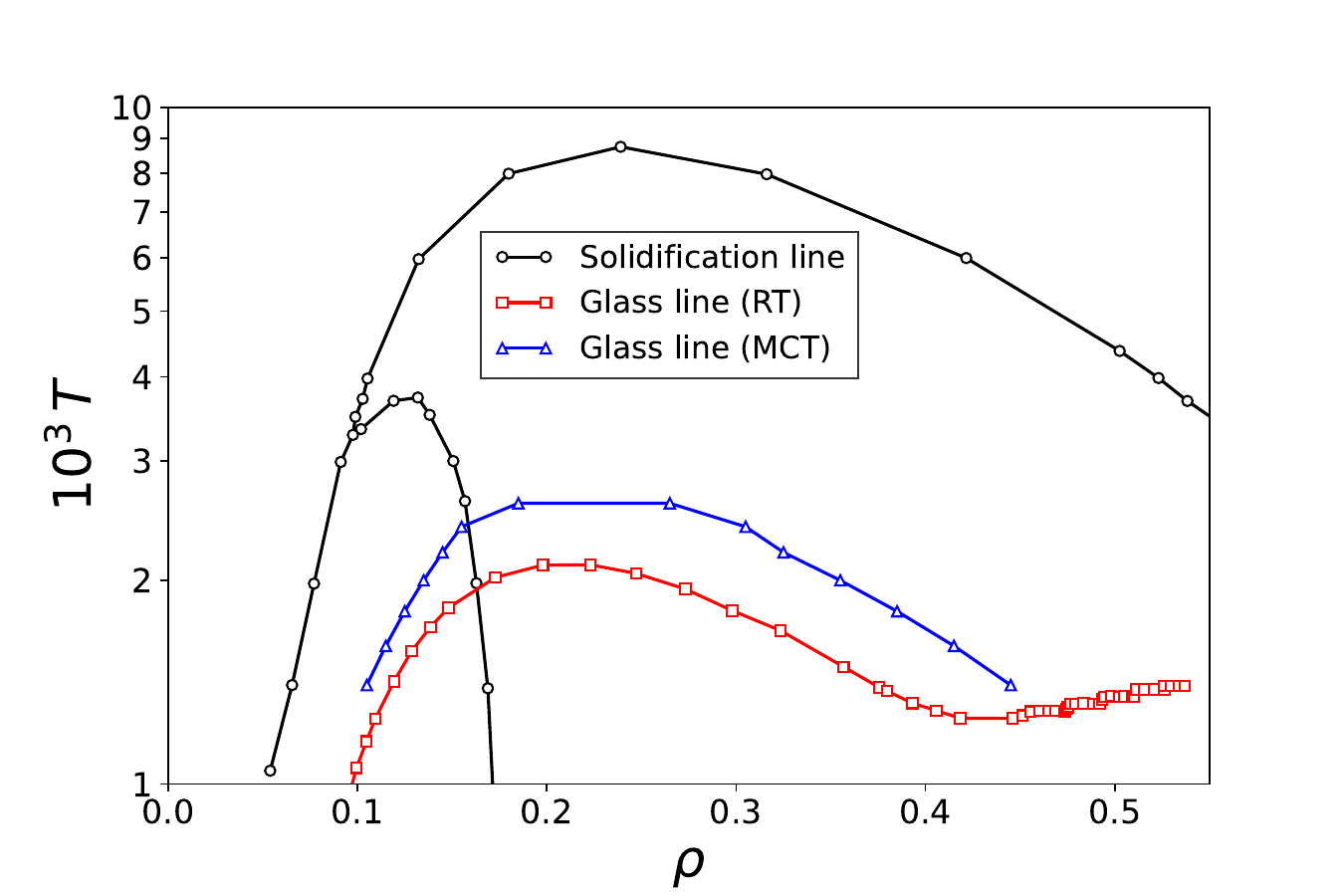}
    \caption{Phase diagram of the GCM. The black solidification line and the red glass line from the Replica Theory (RT) are taken from the literature, in particular from Refs.~\cite{prestipino:pre:2005} and \cite{bomont:pre:2022}, respectively. The blue glass line is calculated from Mode Coupling Theory (MCT) as described in Materials and Methods. The MCT line is recovered at densities $\rho \gtrsim 1.00$, where it follows a monotonically decreasing trend with density~\cite{ikeda:prl:2011, ikeda:jcp:2011}; see main text.
    }
    \label{fig:fig1}
\end{figure}

Within the realm of $Q^{+}$ potentials a prominent role is played by the Gaussian Core Model (GCM) introduced by Stillinger in the 70’s~\cite{stillinger:jcp:1976}. The GCM is one of the simplest models for the description of systems such as polymer or dendrimer solutions~\cite{louis:prl:2000, gotze:jcp:2004} and it entails an inter-particle Gaussian-shaped potential
\begin{equation}
    v(r) = \epsilon \exp [-(r/\sigma)^2],
    \label{eqn:eq1}
\end{equation}
where $\epsilon$ and $\sigma$ are the energy and length scales. We define a reduced density $\rho \to \rho \sigma^3$ and a reduced temperature $T \to k_B T / \epsilon $, where $\rho = N/V$ is the number density, $k_B$ is Boltzmann’s constant, and $T$ the absolute temperature.  We measure lengths in units of $\sigma$ and times in units of  $\sigma\sqrt{m/\epsilon}$, where $m$ is the mass of each particle. Whereas at low temperatures and low densities, the equilibrium properties of the GCM can be described by an effective hard-sphere mapping~\cite{stillinger:jcp:1976}, at high densities a mean-field description sets in, giving rise to re-entrant melting below a threshold upper freezing temperature $T_{u} = 8.74 \times 10^{-3}$, above which it remains fluid at all densities~\cite{stillinger:jcp:1976, lang:jpcm:2000, prestipino:pre:2005}.  
The equilibrium phase diagram of the GCM, along with two predictions for its vitrification
line, arising from two different approximations, is shown in Fig.~\ref{fig:fig1}.
Another prominent member of the $Q^+$-class is the Hertzian potential, $v(r) = \epsilon(1 - r/a)^{5/2}\Theta(1 - r/a)$, which models the effective interaction between elastic spheres of diameter $a$~\cite{pamies:jcp:2009}. The general features of the phase diagram of the Hertzian spheres are similar to those of the GCM, however, it must be emphasized that the former has finite support, as expressed by the Heaviside function $\Theta(1 - r/a)$, whereas the GCM  is nonvanishing for arbitrarily large values of $r$.

At low densities, the GCM can be mapped onto an effective hard sphere potential~\cite{stillinger:jcp:1976}, and its crystallization properties can be understood in such a context. The high-density part of the phase behavior of the GCM is more challenging.  Previous studies~\cite{ikeda:prl:2011, ikeda:jcp:2011} have shown that at high densities nucleation is strongly suppressed and that Mode-Coupling Theory (MCT) provides an accurate description of the structural arrest of the GCM into an amorphous, glassy state. Moreover, it was found that the glass state at high densities displays strong dynamic fluctuations and a nearly Gaussian distribution of single particle displacements, features compatible with a geometric transition~\cite{coslovich:pre:2016}. Such transition refers to a change in the topology of the rugged free energy landscape; in particular, below the transition temperature, the potential energy barriers become much larger than the available thermal energy, and the system is trapped close to local minima of the energy landscape~\cite{grigera:prl:2002}.

The low- and intermediate-density glassy regimes of the GCM remain largely unexplored. Assuming that the low-density vitrification scenario follows the hard-sphere paradigm, it is then particularly interesting to ask the question as to how the vitrification scenario evolves towards the high-density regime and whether dynamically distinct glassy states exist in different density regimes of the GCM.
A recent theoretical study~\cite{bomont:pre:2022} showed an unexpected density dependence of the glassy behavior of GCM particles, see Fig.~\ref{fig:fig1}). Similarly to the equilibrium crystallization behavior, the glass line shows a re-entrance upon increasing the density. However, at moderate densities, the characteristic order parameter at constant density displays sudden jumps when increasing the temperature. This trend suggests a transition between two different glasses, a continuous and a  discretized one. 
In particular, the emergence of a discretized glass has been associated with the formation of \emph{out-of-equilibrium} local aggregates. Indeed, as mentioned above, the GCM is a $Q^{+}$ potential for which no clusters form at equilibrium. In contrast, for ultrasoft particles belonging to the $Q^{\pm}$ class for which cluster formation is an equilibrium phenomenon, the emergence of cluster-glasses has been recently reported~\cite{coslovich:jcp:2012, coslovich:sm:2013, miyazaki:prl:2016, miyazaki:jcp:2019}. 
We note that in Fig.~\ref{fig:fig1}, the MCT-vitrification line stops at $\rho \gtrsim 0.40$ and at about the same density the RT-vitrification line shows non-monotonic behavior with density. The reason for the former is a loss of the HNC solution for the one-component CGM, which is however recovered at much higher densities, $\rho \gtrsim 1.00$. The RT, being a two-component approach, does converge in this region and results in the aforementioned non-monotonic behavior. Whereas it is an open question whether this behavior is connected with the convergence problems of the HNC in that region, the discretized glass predicted by the RT  occurs already at lower densities, and thus the question of whether a distinct arrested state exists there is independent of the HNC-convergence issues.

The goal of this work is to characterize the transition from low- to high-density glass from a dynamic point of view, focusing on the study of the supercooled regime. Indeed, when approaching the glass transition the system enters into a supercooled regime which represents in all respects a precursor of the glass~\cite{berthier:rmp:2011}. Thus, we expect to observe differences between the two states already at the level of supercooled (glassy) dynamics. 
The supercooled regime of canonical glass-formers is usually described in terms of the \emph{caging} mechanism: each particle experiences trapping due to the neighboring particles that effectively create a cage around it; eventually, the fluctuations allow the particle to escape this local cage and move to the next one. The lower the temperature the more difficult it will be for the particle to escape from the cage. In these terms, the glass transition can be thought of as a localization transition. The cage size is related to the average inter-particle distance, which in turn depends on the density of the system. Such a mechanism is accurate for systems characterized by a harshly repulsive inter-particle potential, such as hard-sphere or Lennard-Jones systems~\cite{chaudhuri:prl:2007, pastore:sm:2015}. However, when dealing with bonded potentials, that is with potentials that do not diverge when two particles are at full overlap, and in the presence of long tails, this mechanism can break down. In particular, we show that for the GCM at intermediate densities, the idea of slow dynamics based on the concept of \emph{caging} must be revised. 

\section*{Results}

Following the phase diagram in Fig.~\ref{fig:fig1}, we simulate the glassy dynamics of the GCM at different densities. As mentioned above, at high densities the one-component GCM vitrifies and thus, it is possible to approach the supercooled regime directly, without running into crystallization issues. This is not the case for low and intermediate densities, for which the one-component system would crystallize upon cooling. Therefore, in our simulations, we follow a random pinning procedure and freeze a fraction $f=10\%$ of the particles in order to avoid crystallization and be able to approach the deep supercooled regime.  All results reported below are calculated taking into account the mobile particles only and averaged over at least three different realizations of the pinning disorder (see Materials and Methods for more details on the simulation protocol). 

\begin{figure}[tbhp]
    \centering
    \includegraphics[width=0.9\linewidth]{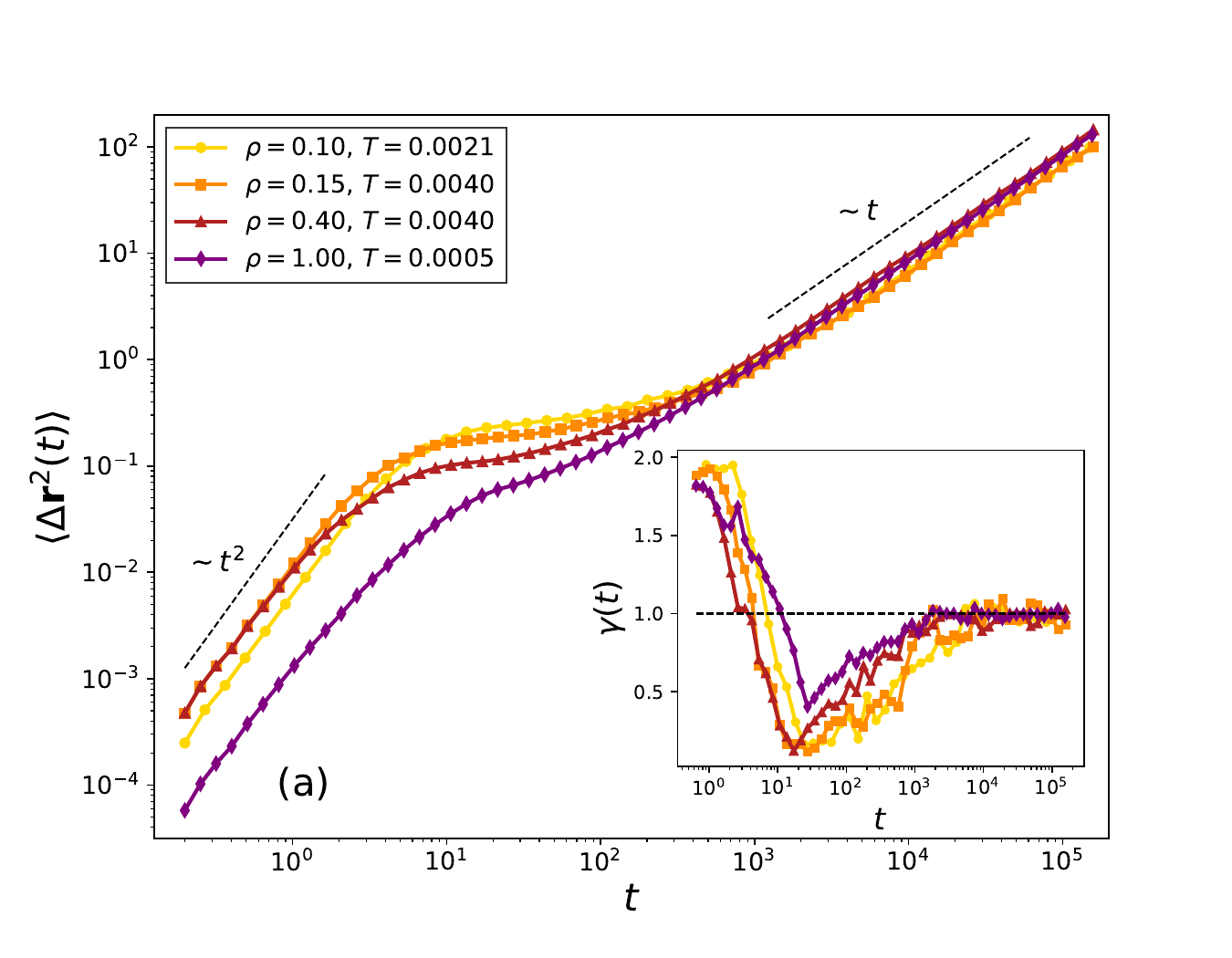}
    \includegraphics[width=0.9\linewidth]{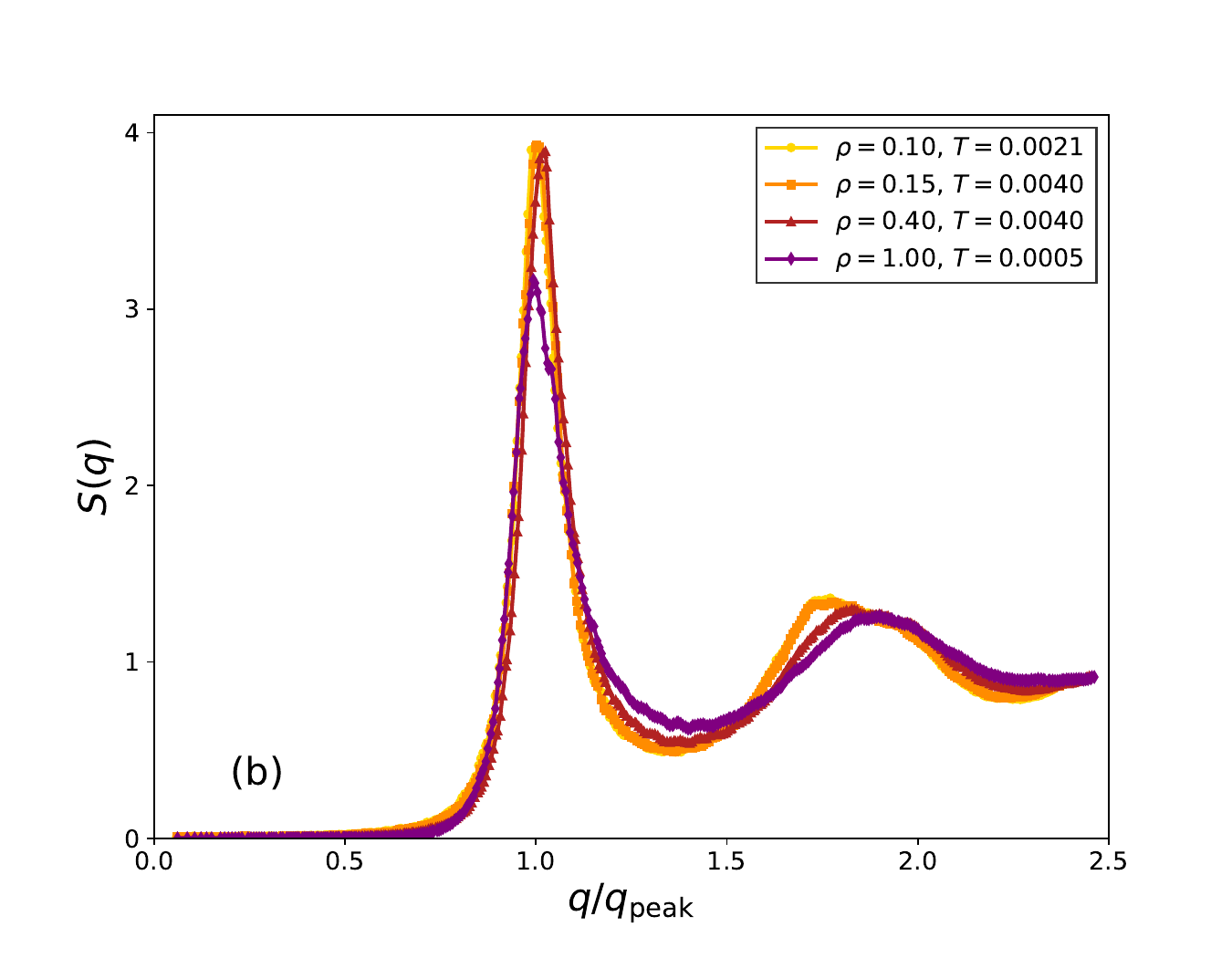}
    \caption{(a) main plot: the mean squared displacement $\langle\Delta\boldsymbol{r}^2(t)\rangle$ of the GCM as a function of time, calculated at four different isodiffusive points, as indicated in the legend. Inset: corresponding local exponent $\gamma(t)$ of the MSD as defined in \eqref{eq:gamma}. (b) structure factors $S(q)$ calculated at the same isodiffusive points as panel (a); in the horizontal axis, the wavenumbers $q$ of each structure factor are rescaled over the value related to the corresponding first peak ($q_\mathrm{peak}$).}
    \label{fig:fig2}
\end{figure}

We focus our analysis on four different densities, \emph{i.e.}, $\rho=0.10, 0.15, 0.40, 1.00$, in order to characterize the transition between low- and high-density glassy dynamics. For each density, we spanned over different temperatures and selected those at which all four systems display the same diffusivity at long times. The motivation behind this choice of isodiffusivity points has different motivations: on the one hand, the isodiffusivity is accompanied by close similarities of the \textit{static} correlations: not only do the iso-peak-height lines of the radial distribution functions follow similar trends as the isodiffusivity ones~\cite{mausbach:fpe:2006, wensink:cpc:2008} but, but upon proper length rescaling, the static structure factors along such lines can be mapped onto a quasi-universal curve as well, as we will shortly demonstrate. On the other, the isodiffusivity lines are precursors of the vitrification line~\cite{foffi:prl:2003}. 
Accordingly, by choosing to simulate along such a locus we are lying on an equidistant line from the 
neighboring dynamically arrested states of the system. To this end, we calculated the mean-square displacements (MSD) as
\begin{equation}
   \langle\Delta\boldsymbol{r}^2(t)\rangle =  \frac{1}{N} \left\langle \sum_{i=1}^N (\boldsymbol{r}_i(t + t_0)- \boldsymbol{r}_i(t_0))^2 \right\rangle,
   \label{eq:msd)}
\end{equation}
where $N$ is the particle number and the brackets $\langle\cdots\rangle$ denote an average over all particles, with initial positions $\boldsymbol{r}_i(t_0)$ at time $t_0$ and $\boldsymbol{r}_i(t + t_0)$ at a time interval $t$ later. We also average over different initial times $t_0$.

The selected temperature values are reported in Fig.~\ref{fig:fig2}(a), in which the mean-squared displacement (MSD) is plotted, clearly showing the same long-time diffusivity for the four systems. We have also calculated the equal-time static structure factor
\begin{equation}
   S(\boldsymbol{q})= \frac{1}{N} 
   \left\langle  \sum_{i=1}^N \sum_{j=1}^N 
   \exp\left[{-{\mathrm{i}}\boldsymbol{q} \cdot \left(\boldsymbol{r}_i(t) -  \boldsymbol{r}_j(t)\right)}\right]\right\rangle,
   \label{eq:sq}
\end{equation}
which is shown in Fig.~\ref{fig:fig2}(b). After suitable rescaling due to the different densities, the overall structural properties of the four systems are comparable. Thermodynamic states along the isodiffusivity lines can be mapped on each other as far as the static correlations are concerned; the  corresponding low- and high-density states lying on an isotherm can be termed \textit{conjuguate pairs}, in analogy to the $T = 0$ pairs of states that are coupled by exact 
duality relations \cite{stillinger:prb:1979}.
This property, together with the same long-time diffusivity, would suggest that a re-mapping of 
all systems is possible and that also the particle dynamics of the associated systems can be collapsed onto a single curve upon suitable rescalings. This, however, is not the case. 

\begin{figure}[tbhp]
    \centering
    \includegraphics[width=1.0\linewidth]{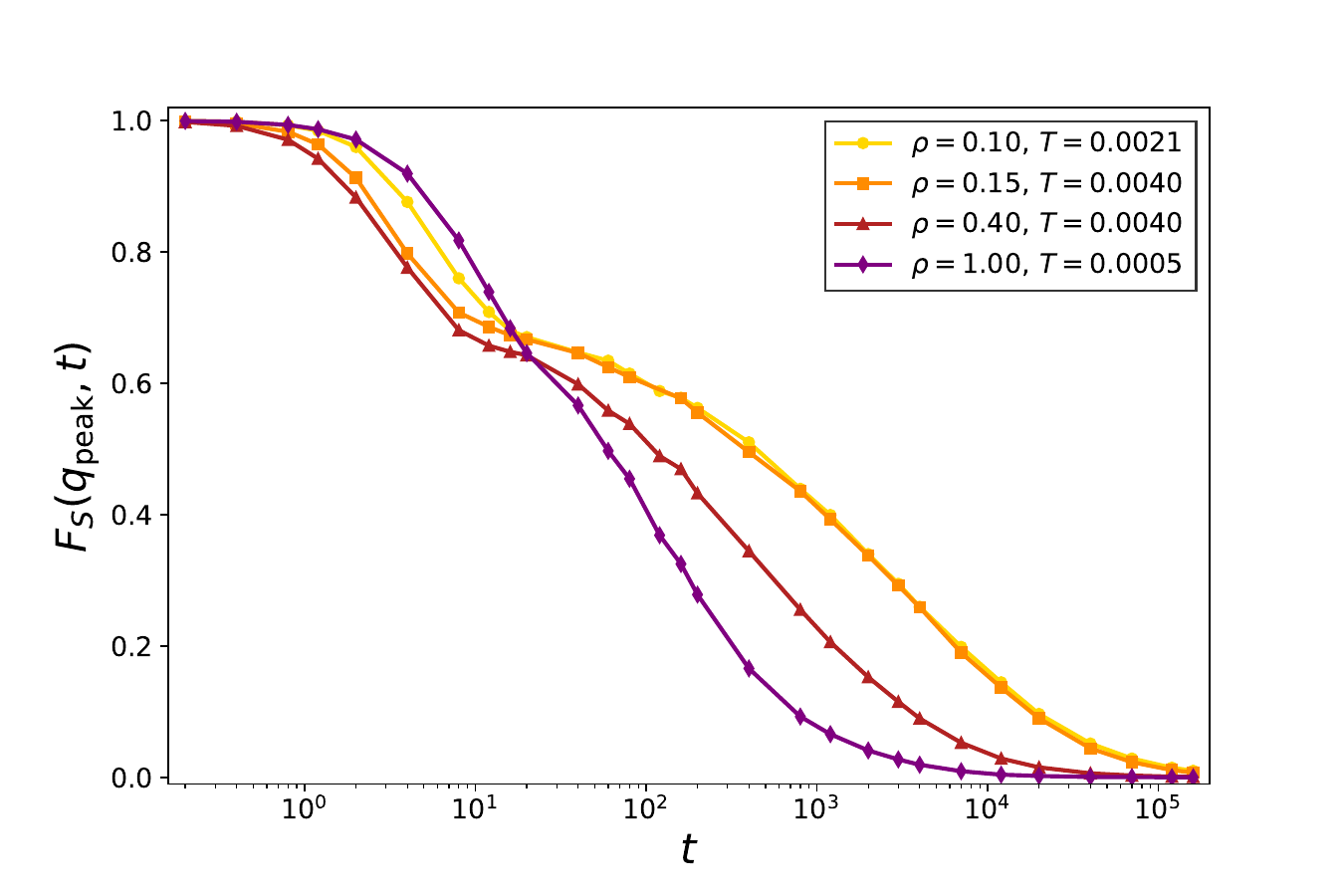}
    \caption{The self-intermediate scattering function 
    $F_s(q_{\mathrm{peak}},t)$ for 
    the four different iso-diffusive points, evaluated at the 
    corresponding wavenumber $q_{\mathrm{peak}}$ where 
    the structure factor of each system has its main peak.
    }
    \label{fig:fig3}
\end{figure}

We consider next the intermediate scattering function (ISF), defined as
\begin{equation}
  F_s(q,t)=\frac{1}{N} \left\langle \sum_{j=1}^N \exp 
  \left[-{\mathrm{i}}\boldsymbol{q} \cdot (\boldsymbol{r}_j(t + t_0)- \boldsymbol{r}_j(t_0))\right]\right\rangle,
  \label{eq:isf}
\end{equation}
which is reported in Fig.~\ref{fig:fig3},  revealing important differences in the relaxation dynamics. In particular, at the two lower densities, a clear two-step relaxation can be discerned, which becomes smoothed out at the intermediate one and practically disappears at the highest of the four, indicating the lack of a caging mechanism for the latter. Therefore, the ISF displays a faster relaxation for high density, while for low and intermediate densities the two-step behavior emerges, leading to an effectively slower relaxation. The coherent part of the ISF displays a similar trend, with no indication of decoupling between self and collective behaviour, at least in the regime investigated within this work. There is, therefore, one single relaxation time associated with both the incoherent and the coherent intermediate scattering functions and not two separate ones, as is the case with other single-component ultrasoft systems, such as semiflexible minirings~\cite{slimani:acsml:2014}.

The aforementioned differences between low- and intermediate densities can already be noticed in Fig.~\ref{fig:fig2}(a) for the behavior of the MSD at times between the ballistic and diffusive regimes. For the lower densities, a much stronger caging effect, indicated by the development of a plateau in the MSD, emerges than for the higher ones. It is worth mentioning that the lack of a plateau at the higher density precisely compensates the fact that the ballistic motion is faster for the lower densities, so that when the former can enter a plateau while the motion at the higher density catches up. Consequently, all MSD's enter their diffusive regime at the same distance squared and at the same time, following thereafter the same diffusive pattern. These differences become quantitative by considering the local exponent $\gamma(t)$ of the MSD, defined as
\begin{equation}
\gamma(t) = \frac{{\mathrm d}\ln \langle \Delta \boldsymbol{r}^2(t)\rangle}{{\mathrm d}\ln t},
\label{eq:gamma}
\end{equation}
and shown in the inset of Fig.~\ref{fig:fig2}(a). Whereas the local exponents for the two lowest densities show a clear transition from ballistic ($\gamma(t) = 2$) to a diffusive ($\gamma(t) = 1$)
regimes through an intermediate plateau ($\gamma(t) \cong 0$), both the ballistic regime and the clear intermediate plateau disappear at the high-density part of the isodiffusivity line. This is a clear indication that at short-to-intermediate scales, the two motions differ, a prediction to be quantified and analyzed below by statistically analyzing individual particle trajectories.

\begin{figure}[tbhp]
    \centering
    \includegraphics[width=1.0\linewidth]{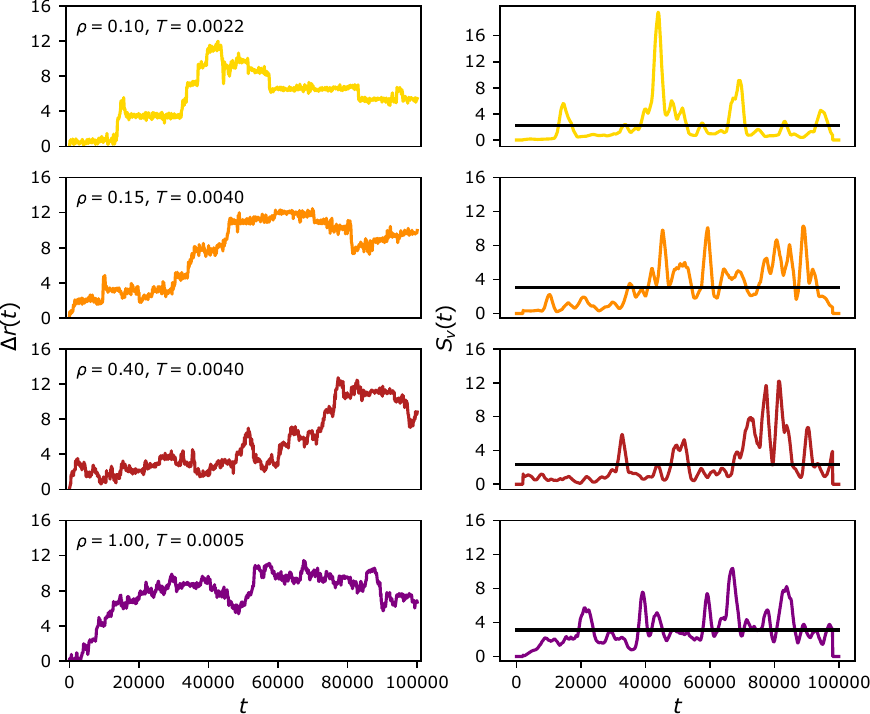}
    \caption{Typical single-particle displacements (left column) and corresponding $S_V(t)$ time series from the LCH analysis (right column) calculated as described in Materials and Methods. The black solid line indicates the threshold dividing slow and fast phases and is calculated as the average over the the whole time series $S_V(t)$, that is $\overline{S_V}$.}
    \label{fig:fig4}
\end{figure}

\begin{figure}[tbhp]
    \centering
    \includegraphics[width=1\linewidth]{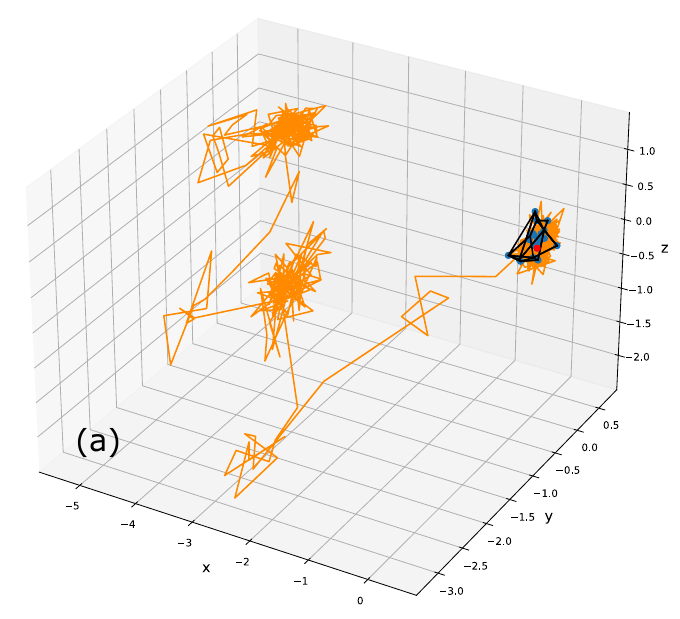}
    \includegraphics[width=1\linewidth]{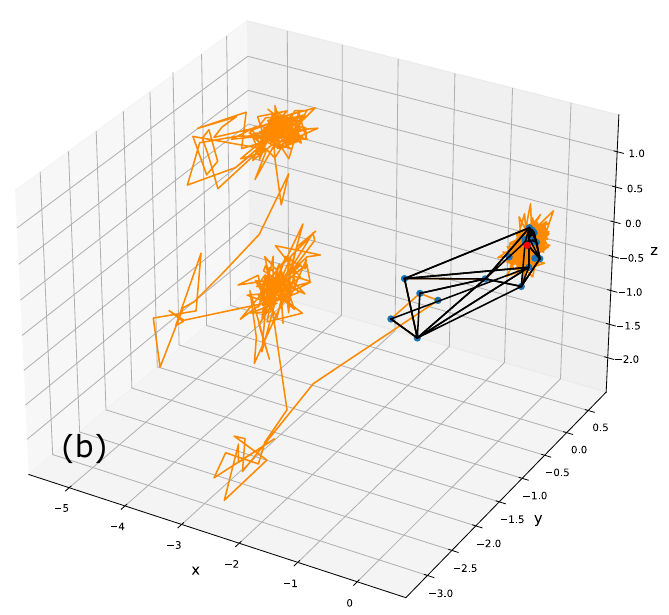}
    \caption{Single-particle trajectory and corresponding LCHs calculated for two sets of points centered in different time instants (red points). The volume of the LCH in (a) is clearly smaller than the one in (b), suggesting that a jump between two local cages can be identified as a peak in the time series of $S_V(t)$ as it is visible in Fig.~\ref{fig:fig4}.
    }
    \label{fig:fig5}
\end{figure}

In Fig.~\ref{fig:fig4} we report typical single-particle displacements, $\Delta r_i(t) = \sqrt{|\boldsymbol{r}_i(t)-\boldsymbol{r}_i(t_0)|^2}$. By looking at the left panels of Fig.~\ref{fig:fig4} it is clear that the dynamics shifts from an intermittent-like behavior at low densities to a more continuous one at high densities. In order to classify this transition we make use of an analysis recently developed to unravel intermittent features in single-particle trajectories and based on a Local Convex Hull (LCH) method~\cite{lanoiselee:pre:2017}. 
The main idea behind this analysis is to use geometric properties of the smallest convex shape (precisely the LCH) enclosing a small set of trajectory points to estimate the space explored by each particle in a specific time window (see Fig.~\ref{fig:fig5}). 
More specifically, our analysis focuses on the study of the LCH volume $S_V(t)$ which, as mentioned in~\cite{lanoiselee:pre:2017}, with respect to other geometric quantities such as the diameter, is more sensitive to changes in the dimensionality and anisotropy of the particle motion. In Fig.~\ref{fig:fig4}, together with the single-particle displacements, we report also the corresponding time series $S_V(t)$ calculated from the LCH method as described in Materials and Methods. We can observe that, if the particle motion has an intermittent-like behaviour, $S_V(t)$ will display few and high peaks, while, if the particle motion follows a more continuous trend, $S_V(t)$ will mostly oscillate around its single-particle average value $\overline{S_V}$ with multiple lower peaks. Such trend suggests that performing a statistical analysis of the $S_V(t)$ peaks can help in classifying different dynamical behaviors. Then, for each time series we find the peak locations and corresponding peak values $S_V^*$. Moreover, we identify with $\Delta t_\mathrm{SP}$ the duration of the so called \emph{slow phases}, that is the time $S_V(t)$ stays below the threshold $\overline{S_V}$ before crossing it. 

\begin{figure}[tbhp]
    \centering
    \includegraphics[width=1.0\linewidth]{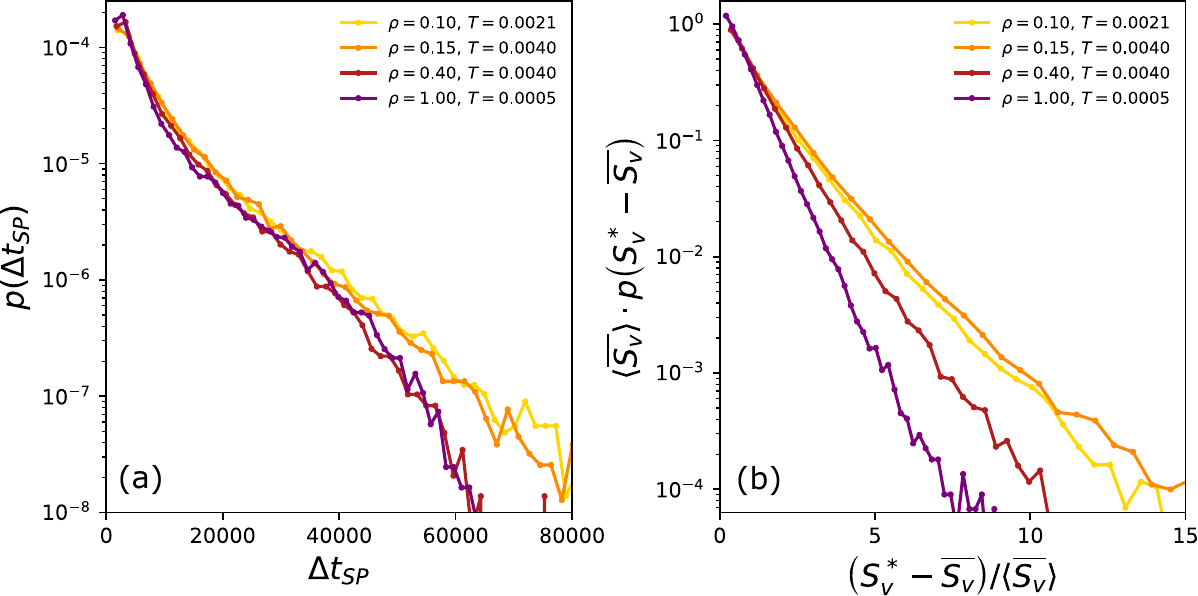}
    \includegraphics[width=0.6\linewidth]{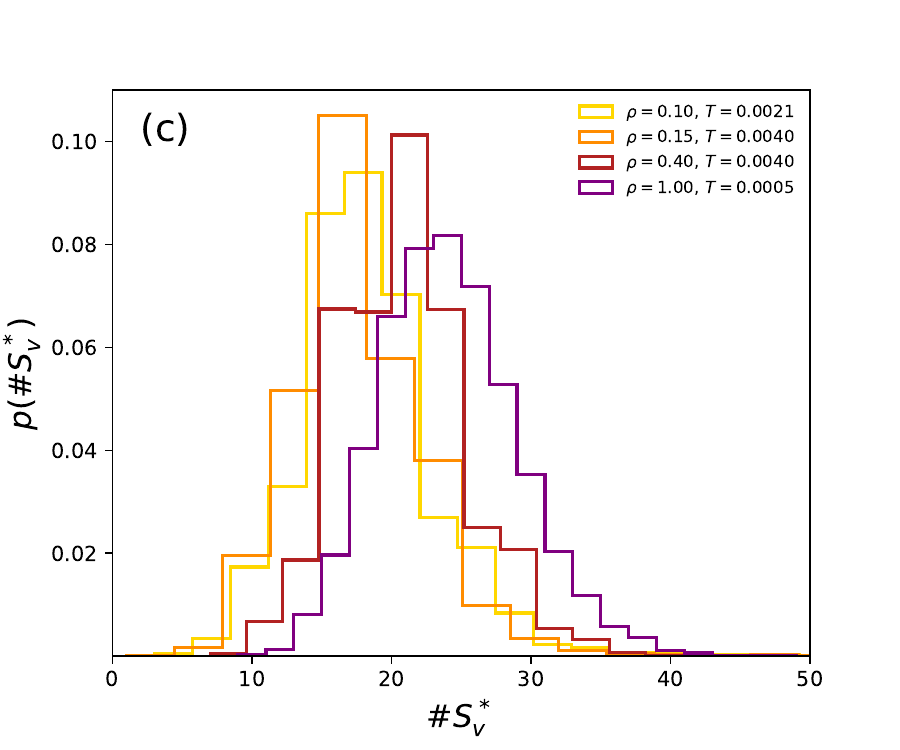}
    \caption{Probability distribution of (a) slow phase duration $\Delta t_\mathrm{SP}$, (b) peak height $S_v^*$ with respect to the threshold value $\overline{S_V}$ and (c) number of peaks. The distribution in (b) is rescaled by the average value $\langle \overline{S_V} \rangle$ which clearly depends on the density (see Fig.~\ref{fig:fig7}).
    }
    \label{fig:fig6}
\end{figure}

\begin{figure}[tbhp]
    \centering
    \includegraphics[width=0.95\linewidth]{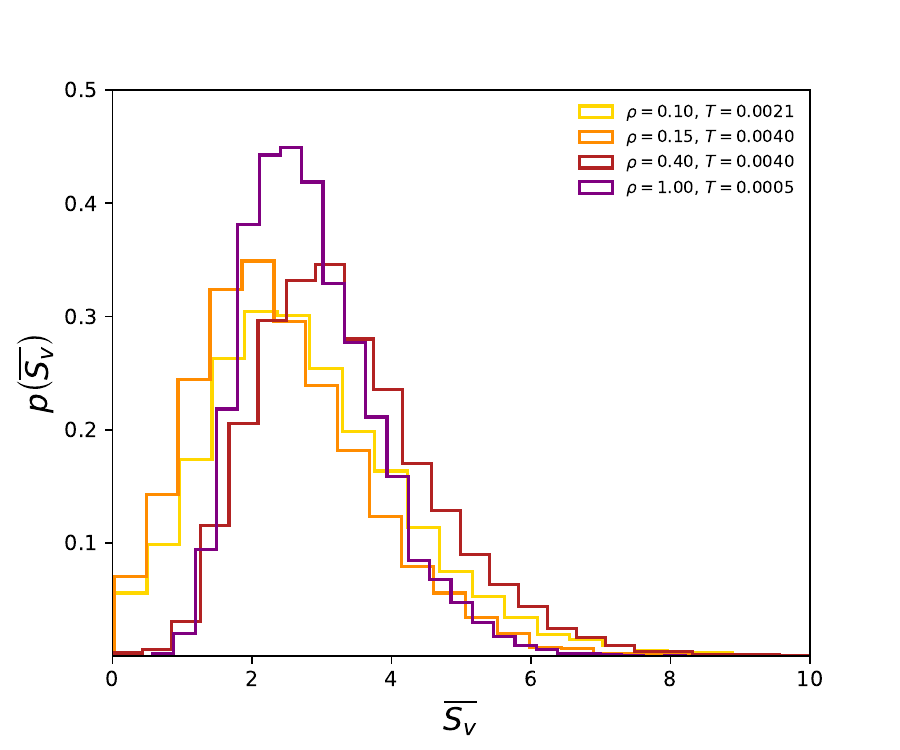}
    \caption{Probability distribution of $\overline{S_V}$ obtained by considering all the values extracted from each single-particle trajectory as indicated in Materials and Methods.}
    \label{fig:fig7}
\end{figure}

In Fig.~\ref{fig:fig6} we report the probability distribution of $\Delta t_\mathrm{SP}$, of the peak height evaluated from the threshold value (\emph{e.g.} $S_V^*-\overline{S_V}$) and of the number of peaks. We can see immediately that $p(\Delta t_\mathrm{SP})$ classifies the systems into two different dynamics: the distribution presents a fatter tail for the two lower density systems with respect to the higher-density systems. In addition, by looking at $p(S_V^*-\overline{S_V})$ we observe that the peak height is more likely to assume larger values for the lower density systems than for the higher density ones. However, in this case, there is not a full rescale of the high-density systems, suggesting that $\rho=0.40$ still belongs to a transition phase between the two dynamical regimes. We emphasize that the quantity $p(S_V^*-\overline{S_V})$ has been rescaled by the average value $\langle \overline{S_V} \rangle$ in order to eliminate trivial contributions stemming from the density-dependence of the volume explored by each particle (see also Fig.~\ref{fig:fig7}).
Finally, the distribution of the number of peaks $p(\#S_V^*)$ complements the information provided by $p(\Delta t_\mathrm{SP})$ suggesting an intermittent-like motion when fewer peaks (and longer slow phases) are detected and a more continuous one when a larger number of peaks (and shorter slow phases) is observed. Indeed, $p(\#S_V^*)$ shows that the system with the highest density is shifted towards larger values with respect to the two systems with lower values, which follow a similar distribution centered around smaller values; once again, the system with $\rho=0.40$ displays an intermediate behavior confirming that at this density the system is in a transition phase between the two dynamical regimes. 
 
We further investigate the particle-to-particle variability of the threshold value $\overline{S_V}$, which indicates the average volume explored by each particle within the simulation time window. To do so, we extract the value $\overline{S_V}$ for each (mobile) particle and then build the histogram, as reported in Fig.~\ref{fig:fig7}. It can be seen that the distribution $p(\overline{S_V})$ looks quite narrow for the highest density, suggesting a more homogeneous dynamics. Conversely, $p(\overline{S_V})$ for the lower density is much broader, implying the presence of slow and fast particles in agreement with the concept of dynamic heterogeneity typical of canonical supercooled liquids. Of particular importance is the fact that in this case, the distribution has contributions even at $\overline{S_V} = 0$, suggesting that, within our time window, there are particles that do not move at all (or only very little) from their initial cage, for the lower densities, whereas this is not the case for the higher densities, for which the distribution vanishes for small values of $\overline{S_V}$.

\begin{figure}[tbhp]
    \centering
    \includegraphics[width=0.49\linewidth]{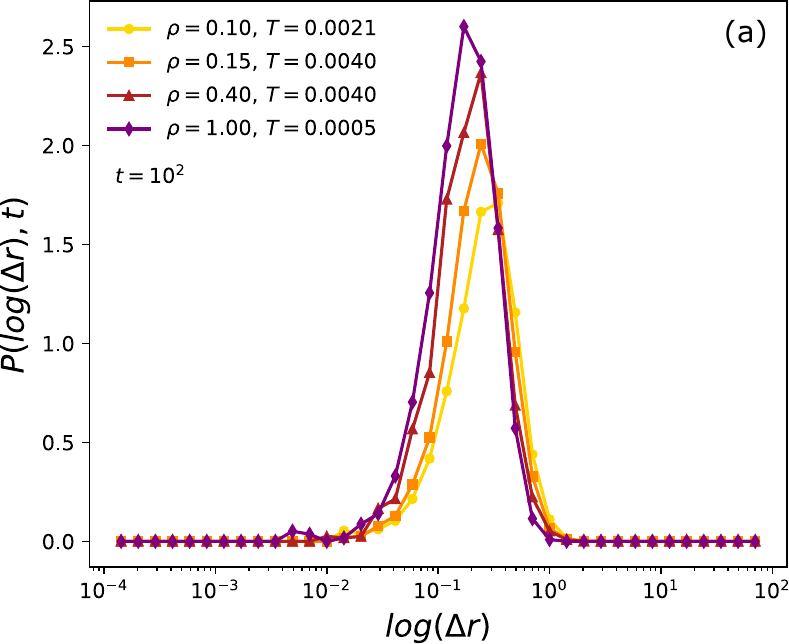}
    \includegraphics[width=0.49\linewidth]{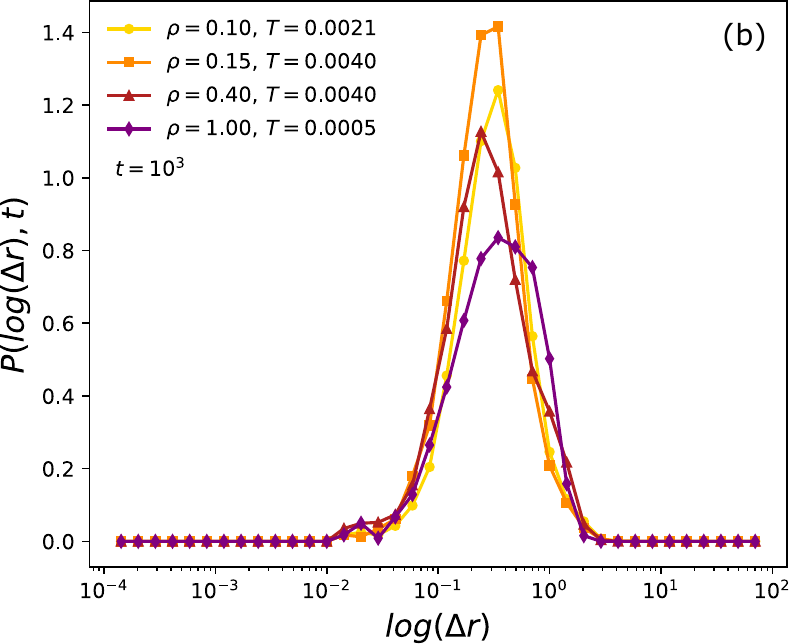}
    \includegraphics[width=0.49\linewidth]{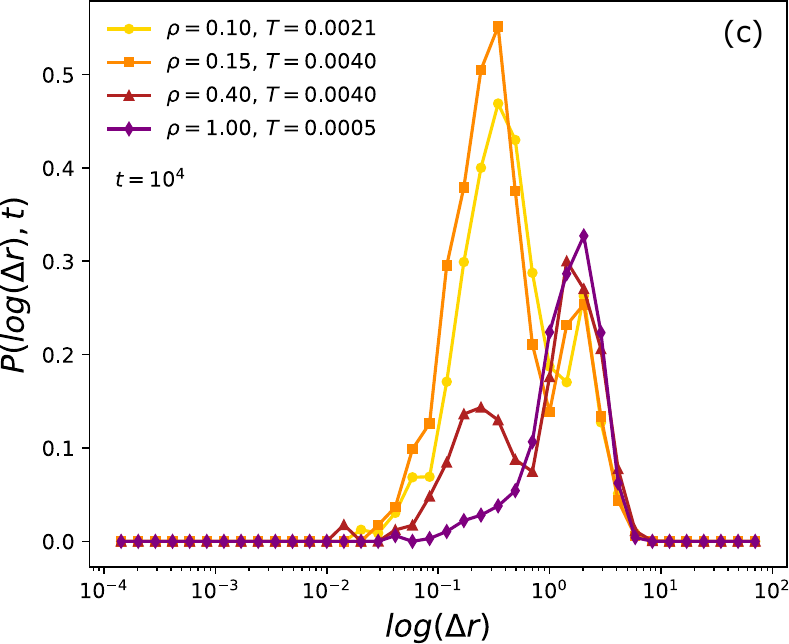}
    \includegraphics[width=0.49\linewidth]{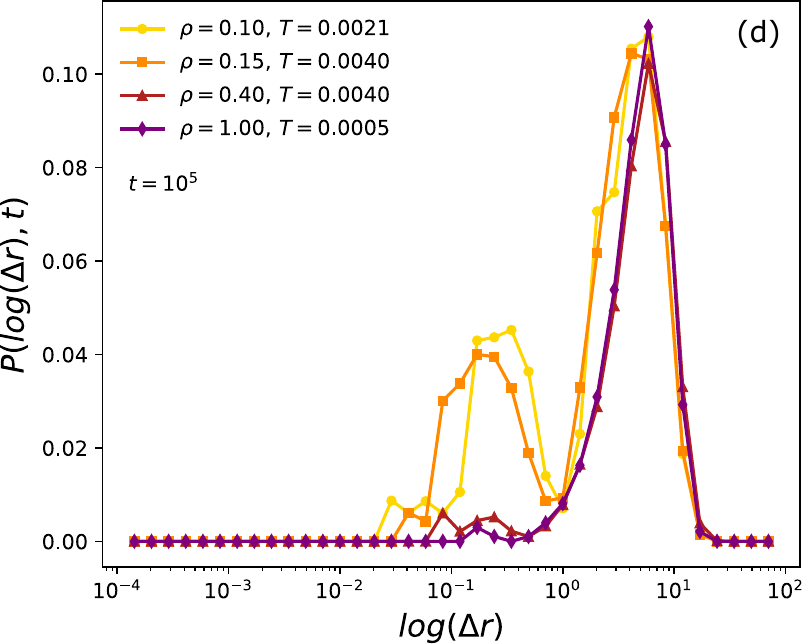}
    \caption{Evolution in time of the single-particle displacement distributions for the four isodiffusivity state points: (a) $t=10^2$, (b) $t=10^3$, (c) $t=10^4$, (d) $t=10^5$. In particular, we calculated the probability distributions of the logarithm of single-particle displacements which allow us to highlight the hopping motion, when present. Indeed, in panel (c) we can clearly see the emergence of a bimodal distribution for all but the highest density system.}
    \label{fig:fig8}
\end{figure}

\begin{figure}[tbhp]
    \centering
    \includegraphics[width=1.0\linewidth]{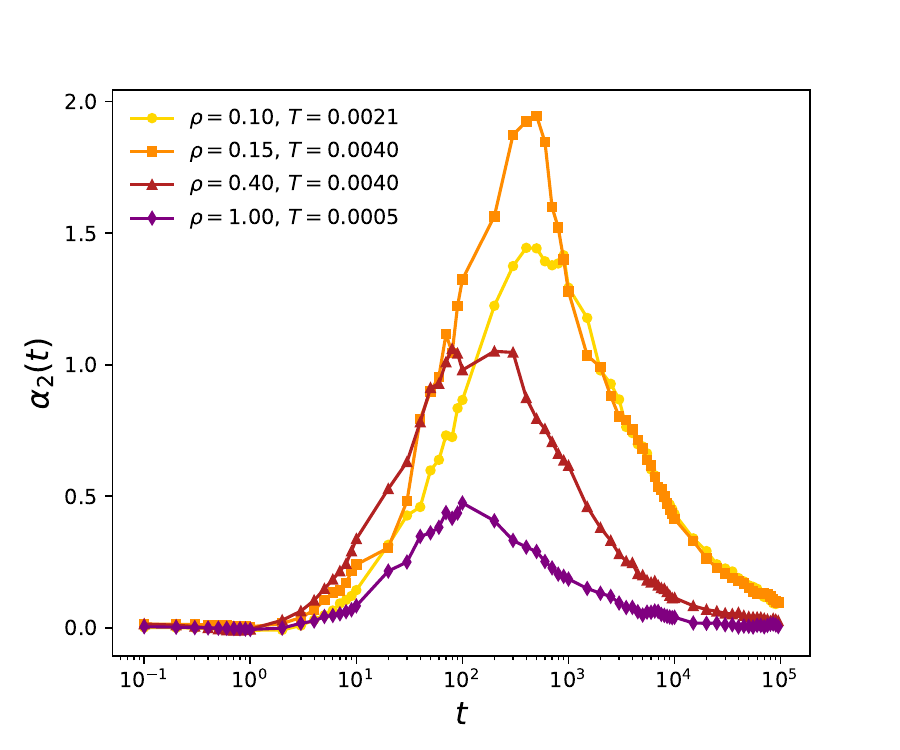}
    \caption{The non-Gaussian parameter $\alpha_2(t)$ of the GCM-supercooled
    fluids at the four isodiffusivity state points considered.}
    \label{fig:fig9}
\end{figure}

Additional corroboration for the gradual disappearing of the standard \textit{cage-escape} mechanism of relaxation for the supercooled  GCM-liquid at intermediate densities is offered by considering the single-particle displacement distribution $P(\Delta r, t)$ and the corresponding non-Gaussian parameter $\alpha_2(t) = 3\langle\Delta r^4(t)\rangle/5[\langle \Delta r^2(t)\rangle]^2-1$, which is used to quantify the deviation of particles displacements from a Gaussian distribution. 
Deviations of this quantity from zero are usually attributed to the presence of dynamic heterogeneity in the system but it has been shown, precisely in the context of the high-density GCM, that small values of $\alpha_2(t)$ are compatible with strong dynamical heterogeneities in the case of a mean-field, geometric glass transition~\cite{coslovich:pre:2016}. 
In Fig.~\ref{fig:fig9} we show the calculated non-Gaussian parameter for the four isodiffusivity state points, finding that those of the low-density points differ drastically from those of the high-density points. In particular, there is a non-monotonic behavior of the curves and of their maximum values, which occurs roughly at the end of the caging time for $\rho = 0.10$ and $\rho = 0.15$ and
somewhat earlier for $\rho = 0.40$ and $\rho = 1.00$.  Thus, $\alpha_2(t)$ follows the same trend in density as the iso-diffusivity line and other quantities characterizing the system, see Fig.~\ref{fig:fig1} and Refs.~\cite{krekelberg:pre:2009,mausbach:fpe:2006,mausbach:zpc:2009}.
Single-particle motions tend thus to become more and more Gaussian as the density grows, in agreement with the absence of a bimodal (cage/cage escape jump) distribution of the self-van Hove function~\cite{coslovich:pre:2016} and thus with the gradual disappearance of the cage-hopping dynamics characteristic of the low-density supercooled GCM fluid. In fact, we explicitly confirm the suppression of hopping dynamics at intermediate and higher densities along our isodiffusivity line in Fig.~\ref{fig:fig8}, demonstrating indeed that the standard caging mechanism of dynamic slowing down in the supercooled liquid is valid only on the low-density side of the vitrification line.

\begin{figure}[tbhp]
    \centering
    \includegraphics[width=1.0\linewidth]{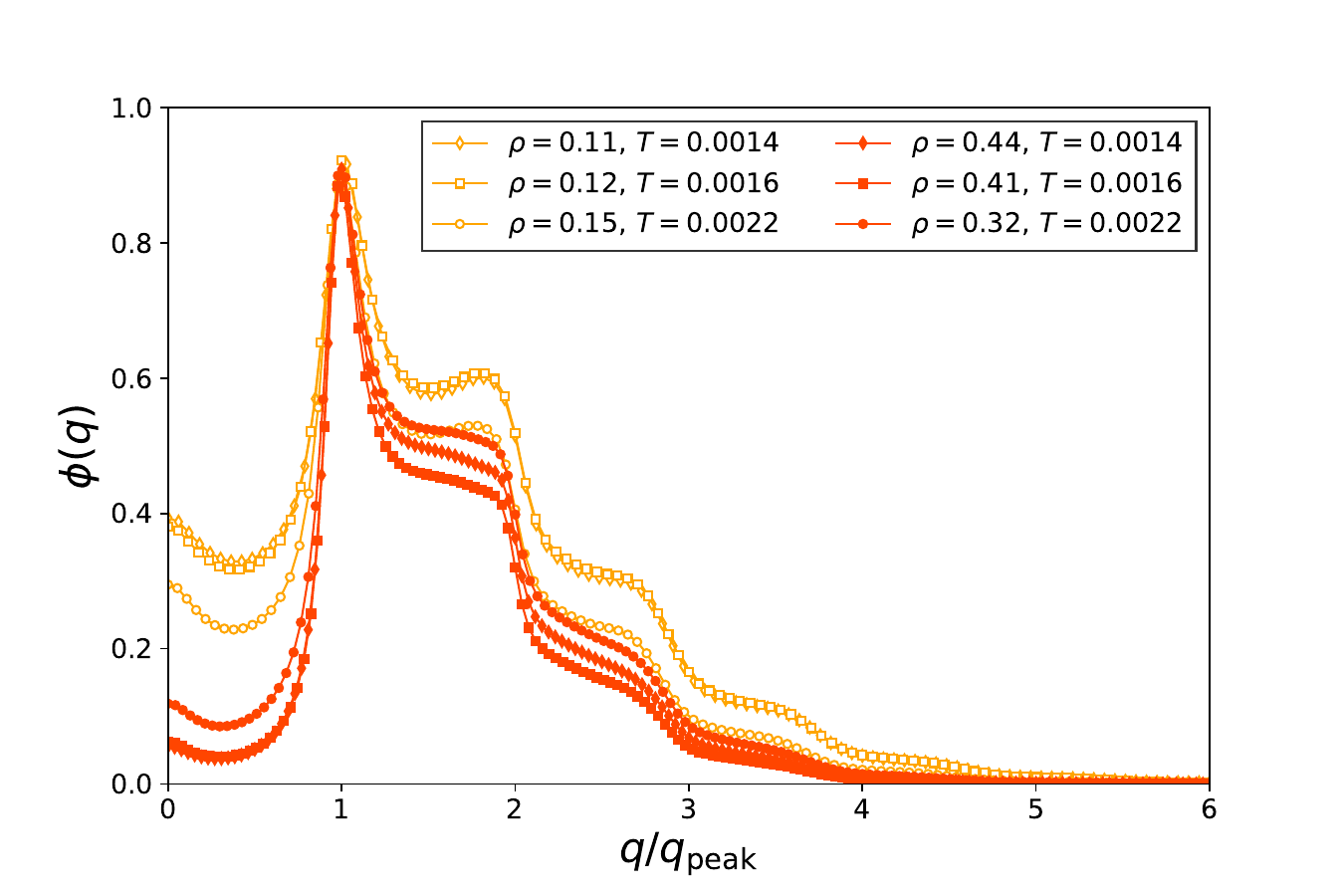}
    \caption{Non-ergodicity factor $\phi(q)$ obtained from MCT (see Materials and Methods). We compare same temperatures at low density (orange curves) and intermediate density (red curves).
    }
    \label{fig:fig10}
\end{figure}

Finally, in Fig.~\ref{fig:fig10} we report the non-ergodicity factor calculated within the MCT framework. As $T$ decreases, the low-density curves tend to behave similarly; the same happens for the high-density graphs. However, comparing at the same temperature,  the non-ergodicity factors show notorious differences, particularly for $q\to0$, suggesting that as $\rho$ grows and the system is getting more and more incompressible, the density modulations of very long wavelengths ($q\to0$) are more and more difficult to become arrested, leading eventually to a different a scenario of dynamic arrest, which is much closer to the ideal MCT-picture and it features long-wavelength \textit{mobility} modulations instead~\cite{ikeda:prl:2011, ikeda:jcp:2011, coslovich:pre:2016}.

\section*{Discussion and Conclusions}

The intermittent dynamics identified for the low-density GCM in the supercooled regime confirms the picture provided by the hard-sphere re-mapping, which appears to remain valid also for the glassy regime. Indeed, in this regime, the GCM displays a glassy behavior that is in agreement with the one of canonical HS-like glass formers. In particular, the dynamics is fully described by the hopping between local cages leading to standard features such as the emergence of the plateau in the MSD, dynamical heterogeneity, and non-Gaussianity. All these features have been re-classified in this work by making use of the LCH analysis. Upon increasing the density, the GCM undergoes a transition towards a different glassy state. The intermittent-like behavior is taken over by a more continuous more Gaussian dynamics. Our analysis shows how the slowing down at high densities of the GCM cannot be explained using the standard caging mechanism. As a matter of fact, the higher the density the less confining the potential becomes due to the increasing contribution of neighbouring particles to the energy landscape. The dynamics in this regime approaches more and more a mean-field-like picture, as already anticipated in~\cite{coslovich:pre:2016}. With our single-particle analysis, we were able to fully capture the transition between these 2 glassy states (HS-like at low densities and mean-field-like at high densities) and to show how the intermediate density regime is characterized by a smooth transition between the two dynamics. Whether this transition
in the supercooled fluid regime 
is the echo of an underlying glass-glass transition
between two arrested states, akin to the distinct glassy states
encountered, e.g., in colloid-polymer mixtures \cite{dawson:pre:2000}, is an
open problem and it will be the subject of further investigations.

\section*{Materials and Methods}

\subsection*{Molecular Dynamics simulations}
Molecular dynamics (MD) simulations were performed using the open-source package LAMMPS ~\cite{thompson2022lammps} for a cubic box with periodic boundary conditions and several combinations of density $\rho = N/V$ and temperature $T$, where $N$ and $V$ are the total number of Gaussian particles and the volume of the box, respectively. The equations of motion were integrated using the velocity Verlet scheme~\cite{frenkelsmit:book}, with a time step $\Delta t/ \tau= 0.01$, where $\tau = \sqrt{m\sigma^2/\epsilon}$ and $m$ the mass unit for the particles. 

Due to the propensity of the system to crystallize at low temperatures, it is required to add some frustration degree into the system, which helps particles to remain in a disordered phase as temperature decreases. In this work, frustration is introduced through random pinning that avoids the inclusion of randomness in the interaction potential or compositional disorder~\cite{berthierkob:pre:2012, karmakar:parisi:pnas:2013}. 

The initial configuration of the system was generated in a multiple steps process: at first, $N_p=\lfloor f N\rfloor$ particles were randomly placed in the simulation box and an equilibration process took place at a relatively high temperature ($T=0.01$). Then they were permanently pinned and the remaining $N_m=N-N_p$ particles (the mobile ones) were randomly inserted in the box preventing excessive overlapping with the previously inserted particles. By setting the target temperature $T$, an equilibration run was performed for the mobile particles in which the system was thermalized by means of the Nose-Hoover thermostat. To consider the effect of the pinning protocol, a second pinning method was also employed. In that case, the whole system is equilibrated at a high temperature, then the fraction $f$ of particles is permanently frozen and the system is lastly quenched to the target temperature for the equilibration of the mobile particles to occur. 

We worked with a total number of particles $N= 3456$,  or $5000$, and $f = 0.05$ or $0.10$. The typical equilibration time for the mobile particles ranged from $2\times 10^6$ to $10^7$ time steps. After reaching a steady state, indicated by the absence of any drift in internal energy and pressure, a production run was performed, which typically ranged from $10^7$ to $10^8$ time steps. 

By providing that all average measurements for given density and temperature are performed over both thermal fluctuations and different realizations of the pinning disorder, it is expected that the thermodynamics of the mobile particles remains unperturbed~\cite{berthierkob:pre:2012}. 


\subsection*{Local convex hull analysis} 

To differentiate between different diffusive phases we analyse the single-particle trajectories making use of the Local Convex Hull (LCH) method developed in~\cite{lanoiselee:pre:2017}. The convex hull of a finite set of points $\{ \mathbf{x}_1, \dots , \mathbf{x}_n\} \subset \mathbb{R}^d $ is the minimal convex shape that encloses all the points  $\mathbf{x}_1, \dots , \mathbf{x}_n $. We consider the volume $Q_V(i)$ of the LCH computed over $2\tau_0 +1$ trajectory points centered around $\mathbf{x}_i$. 
Then, we classify the point $\mathbf{x}_i$ by using all the estimators to which it contributes (in total $2\tau_0+1$), that is by introducing the discriminator:
\begin{equation}
S_V(i) = \frac{1}{2\tau_0 +1} \sum_{k=i-\tau_0}^{k=i+\tau_0}Q_V(k).
\label{eq:discriminator}
\end{equation}
The LCH method consists in: 1) mapping each trajectory into a one-dimensional time series by computing the discriminator $S_V(n)$ for all points $\mathbf{x}_n$; and 2) selecting a threshold $S_V$ such that the points $\mathbf{x}_i$  with $S_V(i)>S_V$ are classified into the \emph{fast} phase whereas the points $\mathbf{x}_i$ with  $S_V(i)\le S_V$ into the \emph{slow} phase. Following Ref.~\cite{lanoiselee:pre:2017}, we set the threshold $S_V$ to be the average value of $S_V(n)$ over the single trajectory,
namely
\begin{equation}
S_V \equiv \overline{S_V} = \frac{1}{N-4\tau_0} \sum_{n=2\tau_0+1}^{n=N-2\tau_0}S_V(n).
\label{eq:svbar}
\end{equation}
The parameter $\tau_0$ was chosen empirically by looking at the MSD curves and selecting the time at which the system escapes from the plateau, generally used as an estimate of the caging time. For our system in particular we have $\tau_0=10^3$ (in simulation time units). To calculate the LCH, we used the available algorithm in Python based on the Qhull library.

\subsection*{Mode Coupling Theory}

To quantify the liquid-to-glass transition of the GCM system, the non-ergodicity factor $\phi(q)$ was evaluated, which is defined as the long-time limit of the  density autocorrelation function, i.e.,
\begin{equation}
\phi(q)=\lim_{t\to\infty}F(q,t)=\lim_{t\to\infty} \frac{\langle\rho({\bf q},t)\rho(-{\bf q},0)\rangle}{\langle\rho({\bf q},0)\rho(-{\bf q},0)\rangle},
\end{equation}
where $\rho({\bf q},t) = \sum_{j=1}^N \exp[{\mathrm{i}}{\bf q}\cdot {\bf r}_j (t)]$ and the sum is performed over the coordinates ${\bf r}_j(t)$ of all particles in the system.

In the case  $\phi(q)\neq0$, the system is considered  non-ergodic and its state is identified as glassy, whereas $\phi(q) = 0$ corresponds to an ergodic fluid.  Given the structural data obtained by solving the Ornstein-Zernike equation through the hypernetted-chain closure (HNC), the calculation of the non-ergodicity factor is readily achieved within the framework of the Mode Coupling Theory (MCT). According to it, $\phi(q)$ fulfills the self-consistent  equation \cite{hansen:book}:
\begin{equation}
    \frac{\phi(q)}{1-\phi(q)} = \frac{1}{(2\pi)^3}\int {\rm d}^3 k\,\mathcal{V}(\mathbf{q},\mathbf{k}) \phi(k) \phi( k^\prime), 
    \label{eq:fk}
\end{equation}
where  $\mathbf{k}^\prime=\mathbf{q}-\mathbf{k}$ and the kernel $\mathcal{V}(\mathbf{q},\mathbf{k})$ can be expressed entirely in terms of the 
Fourier transform of direct correlation function $\hat c(k)$, i.e.,
\begin{equation}
  \mathcal{V}(\mathbf{q},\mathbf{k}) = \frac{\rho S(q)}{2q^4} \left[ (\mathbf{q}\cdot\mathbf{k}^\prime)\,\hat{c}(k^\prime) +(\mathbf{q}\cdot\mathbf{k}) \,\hat{c}(k) \right]^2  S(k) S(k^\prime),
\end{equation}
and $S(k) =\left[1- \rho\,\hat c(k)\right]^{-1}$.

\begin{acknowledgments}
We thank S.~Prestipino for providing the data from Ref.~\cite{prestipino:pre:2005} shown in Fig.~\ref{fig:fig1} and E.~Zaccarelli for helpful discussions. V.S.~acknowledges support of the European Commission through the Marie Sk\l{}odowska-Curie COFUND project REWIRE, Grant Agreement No.~847693. M.C.~thanks  Universidad Antonio Nari\~{n}o for financial support through Project No 2022202.
\end{acknowledgments}



\bibliography{mybib.bib}

\end{document}